\documentclass{SCIS2024}
\usepackage{verbatim}
\usepackage{multirow}
\usepackage{tabularx}
\usepackage{makecell}
\usepackage{amsmath}
\begin{document}
\ArticleType{RESEARCH PAPER}
\SpecialTopic{Special Topic: Integration of Large AI Model and 6G}
\Year{2025}
\Month{}
\Vol{}
\No{}
\DOI{}
\ArtNo{}
\ReceiveDate{}
\ReviseDate{}
\AcceptDate{}
\OnlineDate{}

\title{Large AI Model for Delay-Doppler Domain Channel Prediction in 6G OTFS-Based Vehicular Networks}{LAIM for Channel Prediction}

\author[1]{Jianzhe Xue}{}
\author[1]{Dongcheng Yuan}{}
\author[1]{Zhanxi Ma}{}
\author[1]{Tiankai Jiang}{}
\author[1]{Yu Sun}{}
\author[1]{\\Haibo Zhou}{{haibozhou@nju.edu.cn}}
\author[2]{Xuemin Shen}{}

\AuthorMark{Jianzhe Xue}

\address[1]{School of Electronic Science and Engineering, Nanjing University, Nanjing, {\rm 210023}, China}
\address[1]{Department of Electrical and Computer Engineering, University of Waterloo, Waterloo, {\rm N2L3G1}, Canada.}

\abstract{
Channel prediction is crucial for high-mobility vehicular networks, as it enables the anticipation of future channel conditions and the proactive adjustment of communication strategies. However, achieving accurate vehicular channel prediction is challenging due to significant Doppler effects and rapid channel variations resulting from high-speed vehicle movement and complex propagation environments. In this paper, we propose a novel delay-Doppler (DD) domain channel prediction framework tailored for high-mobility vehicular networks. By transforming the channel representation into the DD domain, we obtain an intuitive, sparse, and stable depiction that closely aligns with the underlying physical propagation processes, effectively reducing the complex vehicular channel to a set of time-series parameters with enhanced predictability. Furthermore, we leverage the large artificial intelligence (AI) model to predict these DD-domain time-series parameters, capitalizing on their advanced ability to model temporal correlations. The zero-shot capability of the pre-trained large AI model facilitates accurate channel predictions without requiring task-specific training, while subsequent fine-tuning on specific vehicular channel data further improves prediction accuracy. Extensive simulation results demonstrate the effectiveness of our DD-domain channel prediction framework and the superior accuracy of the large AI model in predicting time-series channel parameters, thereby highlighting the potential of our approach for robust vehicular communication systems.}


\maketitle

\section{Introduction}

The high mobility and complex environments of vehicular networks pose significant challenges for wireless communication, characterized by doubly-selective signal distortion and rapid fluctuations in channel path loss, which hinder hyper reliable and low latency communications (HRLLC) \cite{6GV2X,SCIS_AI,RAN_Survey,SCIS_Blockchain}. High vehicle speeds cause time-selective fading due to the Doppler effect, while multi-path propagation from obstacles leads to frequency-selective fading. Orthogonal time frequency space (OTFS) modulation has emerged as a promising sixth-generation (6G) technology for vehicular communications by operating in the delay-Doppler (DD) domain, effectively mitigating both types of fading and enhancing communication robustness in dynamic environments \cite{OTFS_first}. However, rapid fluctuations in channel path loss can reduce the signal-to-noise ratio (SNR) below the required thresholds, compromising reliability. Therefore, channel prediction becomes essential, allowing networks to anticipate future conditions and proactively adjust communication parameters to maintain stable and reliable connections \cite{Prediction_GANGRU,Prediction_TimeGAN,WX_edge,ChannelDeduction,DT_Positioning}.

Channel prediction in vehicular networks is exceptionally challenging due to the combination of high vehicle mobility and complex propagation environments \cite{VCP_Realistic}. The rapid movement of vehicles induces significant Doppler effects, resulting in a severe Doppler shift \cite{URLLC}. This Doppler shift leads to a very short channel coherence time, which can be less than 1 millisecond in vehicular scenarios. Additionally, multi-path propagation causes transmitted signals to arrive at the receiver via multiple paths due to reflections from buildings, vehicles, and other obstacles. This results in multiple copies of the signal arriving at slightly different times and with different phases. These copies interfere constructively or destructively, leading to rapid fluctuations in signal amplitude known as fast fading and deep fading. Consequently, the time-frequency (TF) domain channels exhibit doubly-selective fading owing to the combined effects of multi-path propagation and Doppler shifts. Moreover, characterizing each time and frequency point in the TF domain results in an extensive number of channel parameters. This complex and swift variation of the TF domain vehicular channel makes accurate channel prediction exceedingly difficult \cite{ChannelModels6G}.

To this end, we leverage the DD domain based on OTFS modulation to provide an intuitive, sparse, and stable channel representation, which is particularly beneficial for channel prediction in vehicular networks \cite{TFtoDD}. Firstly, the DD domain channel representation offers an intuitive depiction of the wireless channel by characterizing it through channel response coefficients, propagation delays, and Doppler shifts of each sub-path separately. These parameters directly reflect the underlying physical phenomena of signal propagation, facilitating a clearer understanding and more accurate modeling of signal behavior as vehicles move \cite{OTFS_Prediction}. Secondly, the channel representation in the DD domain is inherently sparse, as only a few significant sub-paths contribute meaningfully to the received signal \cite{OTFS_ISAC}. This sparsity reduces the number of parameters that need to be estimated and predicted, in contrast to the dense grid of coefficients typical of TF domain representations. Thirdly, the DD domain offers a stable representation over time, even under rapid vehicular movement \cite{OTFS_CC}. While TF domain coefficients may fluctuate rapidly due to high mobility and dynamic environments, the DD domain parameters tend to vary more gradually. This relative stability enhances the predictability of the channel. Therefore, with the DD domain channel representation, the complex vehicular channel variations can be transformed into variations of a set of time series parameters with good predictability in the DD domain. Consequently, channel prediction in vehicular networks becomes the prediction of these time-series DD domain parameters.

To accurately predict time series parameters in the DD domain, we employ the large artificial intelligence (AI) model, specifically transformer-based neural networks with billions of parameters, for channel prediction in vehicular networks. The large AI model offers significant advantages over small models, such as long short-term memory (LSTM) networks, by effectively capturing the complex temporal correlations inherent in time series data and enabling more precise modeling of intricate channel variation patterns \cite{LLMTimeSeries,Timer,Edge_LLM, Network_LLM}.
A notable benefit of the large AI model is its zero-shot prediction capability, allowing it to be cost-effectively and flexibly deployed without the need for specialized training on vehicular channel data \cite{ZeroShot}. Pre-trained on vast amounts of general time series data, these models develop a comprehensive understanding of temporal dynamics that can be applied to unseen channel conditions, thereby allowing them to generalize effectively and make accurate predictions even in new environments.
Furthermore, the large AI model can be fine-tuned with actual vehicular channel data to further enhance their prediction accuracy, adapting their learned representations to the specific characteristics of channel data \cite{FineTuning}. This makes the large AI model particularly powerful for addressing the challenges of channel prediction in high-mobility vehicular networks, outperforming smaller models in both accuracy and robustness.

In this paper, we propose a DD domain channel prediction framework for high-mobility vehicular networks, and employ the large AI model to predict the time-series parameters inherent in the DD domain channel representation. Specifically, in OTFS-based vehicular networks, channel estimation is naturally performed in the DD domain, providing a representation where channel parameters vary continuously and stably over time with excellent predictability. We transform the vehicular channel prediction problem into a time series prediction task for DD domain parameters, including the response coefficient, propagation delay, and Doppler shift of each sub-path. This significantly reduces the complexity associated with doubly-selective vehicular channel prediction. Additionally, by employing the large AI model, we achieve accurate and robust predictions of DD domain channel parameters. The zero-shot capability of the large AI model allows for low-cost and flexible channel prediction deployment without prior specialized training, while fine-tuning with specific vehicular channel data further enhances prediction accuracy. The combination of the stability of DD domain channel representations and the advanced predictive capability of the large-scale AI model enables our approach to effectively address the challenges posed by high mobility and complex propagation environments of vehicular networks. Besides, we demonstrate the effectiveness of our channel prediction approach through extensive simulations. Three main contributions of this paper are summarized as follows: 

\begin{itemize}
\item We utilize OTFS modulation in vehicular networks to effectively counteract
the signal distortions caused by doubly-selective channels. Meanwhile,
OTFS modulation introduces the DD domain, offering a stable, sparse,
and intuitive representation of the vehicular channel that aligns
closely with the physical propagation environment.
\item We propose a DD domain channel prediction framework tailored for vehicular
networks characterized by high mobility and complex environments.
Leveraging the DD domain channel representation, channel prediction
is transformed into a time series prediction task of DD domain parameters,
which vary continuously and predictably over time.
\item We leverage the large AI model to accurately and robustly predict time-series DD domain channel parameters, capitalizing on their advanced capacity to model temporal correlations. Their zero-shot capability facilitates low-cost deployment of channel prediction, while subsequent fine-tuning further enhances prediction accuracy. Compared to traditional DL approaches, the large AI model offers a flexible, accurate, and cost-effective solution for vehicular channel prediction.
\end{itemize}

The remainder of this paper is organized as follows. Section 2 reviews
related work on channel prediction. Sections 3 and 4 introduce the
vehicular channel model and the DD domain channel representation,
respectively, and formulate the DD domain channel prediction problem.
Section 5 presents the large AI model used for channel prediction.
Experimental results are given in Section 6, and conclusions are
drawn in Section 7.

\section{Related work}

Channel prediction plays a crucial role in enhancing vehicular network
performance, with numerous studies focusing on this area. Unlike traditional
static channel prediction, vehicular networks require methods that
address the rapid and complex changes in vehicular channels caused
by vehicle movements. The channel prediction problem in high-mobility
massive multiple input multiple output (MIMO) system is investigated
in angle-delay domain, and a spatio-temporal autoregressive
model-based unsupervised-learning method is proposed \cite{VCP_highmobility}.
To cope with rapid channel variation, based on the estimated mobility parameters, a channel prediction approach incorporating the mobility of both scatterers and transceivers is
proposed \cite{VCP_MobilityInduced}. A model-driven
high-accuracy and low-complexity channel state information (CSI) prediction
scheme is presented by pre-estimating the parameter range with the short-time Fourier transform and obtaining the initial values of the parameters with the polynomial Fourier transform \cite{VCP_PFT}.

Deep learning (DL) models with their powerful spatial-temporal data mining
capabilities have shown great promise in improving channel prediction
accuracy by capturing spatial and temporal correlations \cite{CP_NN,TGASA,STDatamining,STDataCenter,Image_MM,PrivacyPrediction}.
A DL channel prediction approach using LSTM networks
is proposed to enhance CSI forecasting in edge computing networks
for intelligent connected vehicles, which outperforms conventional
models like autoregressive integrated moving average model through
extensive simulations \cite{VCP_LSTM}. Gruformer achieves accurate
channel prediction while making training more stable by integrating
the gate recurrent unit (GRU) module into the transformer architecture
\cite{VCP_Gruformer}. A transformer-based parallel channel prediction
scheme is proposed to address the channel aging problem in mobile environments,
enabling accurate predictions across multiple frames simultaneously
and mitigating error propagation, thus achieving negligible performance
loss in practical communication systems under mobile environments \cite{VCP_Transformer}.

Inspired by the success of generative pre-trained transformer (GPT) in
natural language processing, some researchers have begun exploring
whether large AI models could enhance channel prediction performance,
given that time series channel data can be similarly represented as
sequential data \cite{LLMCom,CP_ChannelGPT,TimeLLM,BigAIModel6G,BeamPrediction,GAI_Vehicle,LOSEC}. LLM4CP is a pre-trained large language model (LLM) empowered method for predicting future downlink CSI based on historical
uplink data, utilizing tailored modules for cross-modality knowledge
transfer, achieving high accuracy with low training and inference
costs in massive MIMO systems \cite{CP_LLM4CP}. CSI-GPT integrates
a swin transformer-based channel acquisition network with a variational
auto-encoder-based channel sample generator and federated-tuning to
efficiently acquire downlink CSI in massive MIMO systems \cite{CP_CSIGPT}.
CSI-LLM is an LLM-based channel prediction approach, which
models the historical variable-step sequences, utilizing the next-token generation ability of LLM to predict the CSI of the next step \cite{CP_CSILLM}. However, these studies focus primarily on channel
prediction in static environments. To the best of our knowledge, this
work is the first to apply the large AI model for channel prediction
in high-mobility environments, specifically in scenarios like vehicular networks.

\section{System model}

In vehicular networks, multi-path propagation leads to frequency-selective fading as signals arrive via multiple sub-paths with different delays. While channels in static scenarios that only contain the multi-path effect can be approximated as the linear time-invariant system, vehicle mobility introduces Doppler shifts, making the channel time-variant. Consequently, vehicular channels exhibit both frequency-selective and time-selective fading, resulting in rapid channel fluctuations. Accurately modeling vehicular network channels necessitates capturing both multi-path and Doppler effects arising from continuous vehicle movement and complex propagation environments. High spatial-temporal continuity is essential to represent the rapid channel variations caused by constantly changing vehicle positions and speeds. Statistical channel models can efficiently consider both multi-path and Doppler effects with low computation cost, but they lack the spatial-temporal continuity required for high-mobility scenarios. Ray-tracing channel models provide detailed channel simulations, but they are computationally demanding and challenging in achieving high spatial-temporal resolution.

To accurately model both multi-path and Doppler effects with high spatial-temporal continuity and resolution, we introduce the quasi-deterministic channel model. This channel model generates the propagation environment through a statistical distribution and then allows transmitters and receivers to interact with the environment in a deterministic way. In this section, this channel model is introduced in terms of large-scale fading and small-scale fading.

\subsection{Large-scale fading}

Large-scale fading describes the average signal attenuation over long distances and is characterized by parameters such as the Rician K-factor, shadow fading, and angular domain parameters. Fluctuations in these parameters result from environmental conditions and changes in the relative positions of the transmitter and receiver. Achieving smooth transitions in large-scale parameters and good spatial consistency is crucial for accurate large-scale fading modeling in time-varying channels. 

Specifically, the Rician K-factor quantifies the ratio of direct sub-path gain to scattered sub-path gain, reflecting the relative strengths of line-of-sight (LOS) and non-line-of-sight (NLOS) components. The spatially correlated Rician K-factor \(\tilde{R_{\mathrm{K}}}\) is calculated as, 
\begin{equation}
\begin{aligned}
\tilde{R_{\mathrm{K}}}(\mathbf{p})={R_{\mathrm{K}}}_{\mu}+{R_{\mathrm{K}}}_{\sigma} \cdot \tilde{X}^{R_{\mathrm{K}}}(\mathbf{p}), 
\end{aligned}
\end{equation}

\noindent where \({\tilde{X}^{R_{\mathrm{K}}}}\) is the spatially correlated random variable, \(\mathbf{p}\) is position vector, \({R_{\mathrm{K}}}_{\mu}\) is the expectation of reference value, and \({R_{\mathrm{K}}}_{\sigma}\) is the standard deviation of reference value. The \({R_{\mathrm{K}}}_{\mu}\) is calculated as, 
\begin{equation}
\begin{aligned}
R_{\mathrm{K}}^{\mu} = R_{\mathrm{K}}^{0} +
R_{\mathrm{K}}^{\mathrm{f}} \log_{10}\left(f_{\mathrm{G}}\right) +
R_{\mathrm{K}}^{\mathrm{d}} \log_{10}\left(d_{2 \mathrm{D}}\right) +
R_{\mathrm{K}}^{\mathrm{h}} \log_{10}\left(h_{B}\right) +
R_{\mathrm{K}}^{\alpha} \alpha_{\mathrm{R}} ,
\end{aligned}
\end{equation}

\noindent where \(R_{\mathrm{K}}^{\mathrm{f}}\), \(R_{\mathrm{K}}^{\mathrm{d}}\), \(R_{\mathrm{K}}^{\mathrm{h}}\), and \(R_{\mathrm{K}}^{\alpha}\) represent the impacts of dependent variables in scenarios, including carrier frequency \(f_{\mathrm{G}}\)(Hz), 2D horizontal distance \(d_{\mathrm{2D}}\)(m), base station height \(h_{B}\)(m), and elevation angle seen from the RX \(\alpha_{\mathrm{R}}\)(rad). And \(R_{\mathrm{K}}^{0}\) is the baseline value of the parameter, providing a foundational level for Rician K-factor under specific environmental conditions. The impact of each parameter varies according to statistical distributions specific to different environments.
The spatially correlated random variable \(\tilde{X}^{R_{\mathrm{K}}}\) is calculated as,
\begin{equation}
\begin{aligned}
\begin{array}{l}
\tilde{X}^{R_{\mathrm{K}}}\left(\mathbf{p}\right) 
=\frac{\tilde{X}^{R_{\mathrm{K}}}\left(\mathbf{p}_{tx}\right)+\tilde{X}^{R_{\mathrm{K}}}\left(\mathbf{p}_{rx}\right)}{2 \cdot \sqrt{\rho\left(d\right)}+1} \\
\end{array},
\end{aligned}
\end{equation}

\noindent where \({p}_{tx}\) and \({p}_{tx}\) mean the position vector of transmitter and receiver.
\(\rho(d)\) is the autocorrelation function defined as, 
\begin{equation}
\begin{aligned}
\rho(d) &=
\begin{cases} 
\exp\left(-\frac{d^2}{d_{\lambda}^2}\right), & \text{for } d < d_{\lambda} \\
\exp\left(-\frac{d}{d_{\lambda}}\right), & \text{for } d \geq d_{\lambda}
\end{cases}, 
\end{aligned}
\end{equation}

\noindent where \(d_{\lambda}\) is the decorrelation distance threshold. When the real distances \(d\) is smaller than \(d_{\lambda}\), the correlation decays following a Gaussian distribution, otherwise, it decays according to an exponential distribution. This procedure can be similarly applied to other large-scale parameters, such as shadow fading, generating reference values, and spatially correlated variables in the same manner.

Additionally, all large-scale parameters are cross-correlated, as environmental factors often influence multiple parameters simultaneously. The inter-parameter correlation is determined by a covariance matrix, denoted by a positive definite matrix \(\mathbf{R}\) as,

\begin{equation}
\begin{aligned}
\mathbf{R}=\left(\begin{array}{ccc}
1 & \rho_{{R_{\mathrm{K}}}, {F_{\mathrm{S}}}} \\
\rho_{F_{\mathrm{S}}, {R_{\mathrm{K}}}} & 1
\end{array}\right) ,
\end{aligned}
\end{equation}

\noindent where \(F_{\mathrm{S}}\) is the shadow fading, and \(\rho_{F_{\mathrm{S}}, {R_{\mathrm{K}}}}\) is the cross-correlation between the shadow fading \(F_{\mathrm{S}}\) and the Rician K-factor \({R_{\mathrm{K}}}\).
This process ensures the authenticity and spatial consistency of large-scale fading in dynamic vehicular network environments.

Besides, large-scale general path loss \(L_{\mathrm{P}}\) quantifies the reduction in signal gain due to distance-based attenuation. Unlike other large-scale parameters, it lacks spatial correlation and uniformly affects all sub-paths, independent of environmental changes. It provides a baseline reference for average signal attenuation and is calculated as,  
\begin{equation}
\begin{aligned}
L_{\mathrm{P}} = L_{\mathrm{P}}^{0} + 
L_{\mathrm{P}}^{\mathrm{f}} \log_{10}\left(f_{\mathrm{GHz}}\right) +
L_{\mathrm{P}}^{\mathrm{d}} \log_{10}\left(d_{2\mathrm{D}}\right) + 
L_{\mathrm{P}}^{\mathrm{h}} \log_{10}\left(h_{B}\right) + 
L_{\mathrm{P}}^{\alpha} \alpha_{\mathrm{R}},
\end{aligned}
\end{equation}

\noindent where \(L_{\mathrm{P}}^{\mathrm{f}}\), \(L_{\mathrm{P}}^{\mathrm{d}}\), \(L_{\mathrm{P}}^{\mathrm{h}}\), and \(L_{\mathrm{P}}^{\alpha}\) represent the impacts of frequency, horizontal distance, base station height, and elevation angle, respectively. Also, \(L_{\mathrm{P}}^{0}\) is the baseline path loss, providing a foundational level for average signal attenuation.

\subsection{Small-scale fading}

\begin{figure}[t]
\centerline{\includegraphics[width=6in, keepaspectratio]{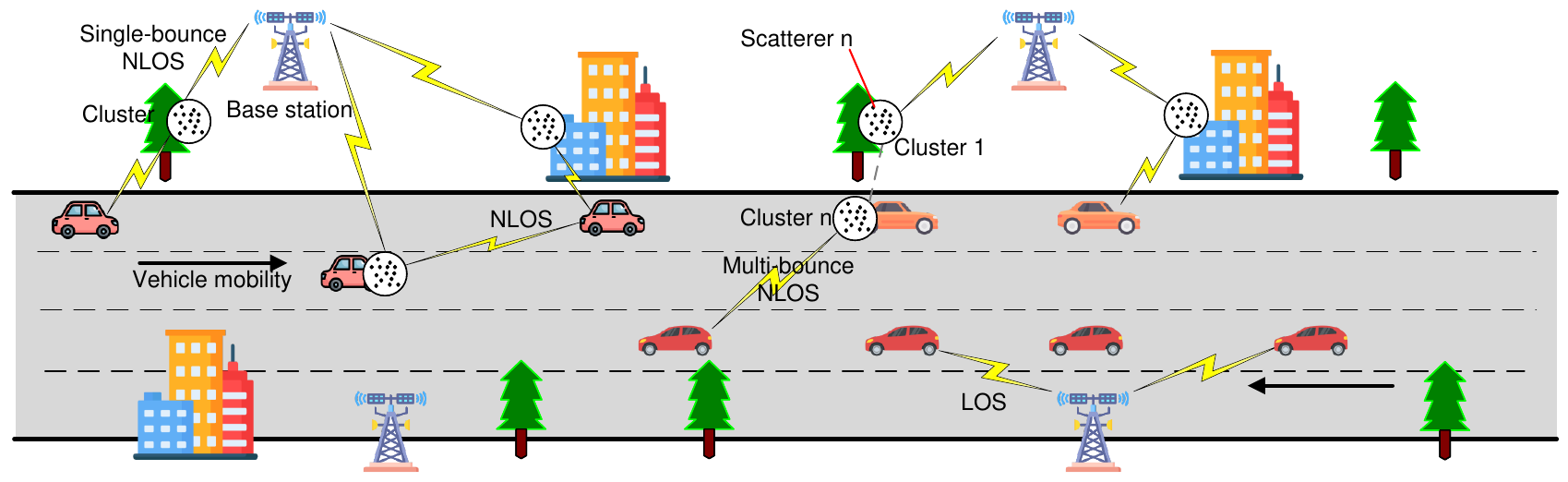}}
\caption{Quasi deterministic channel modeling for high-mobility vehicular networks.}
\label{channelmodel}
\end{figure}

Small-scale fading is modeled using geometric models that capture channel variations at the wavelength level, including rapid time-varying fading caused by multi-path effects and Doppler shifts. In these models, large obstacles like buildings and trees are represented as clusters, which are groups of scatterers that affect the signal path, leading to phenomena such as path diffraction and reflection. 
As shown in Fig. \ref{channelmodel}, the propagation signal may pass through zero or multiple clusters. For multiple sub-paths, each sub-path \(i\) is characterized by variables [\(\tau_{i}\), \(P_{i}\), \(\psi_{i}\)], where \(\tau_{i}\) is the propagation delay, \(P_{i}\) is the gain, and \(\psi_{i}\) is the phase. The channel model updates the signal propagation path in real time based on the movements of the transmitter, receiver, and clusters.

To accurately represent signal strength and phase variations due to antenna directionality and polarization in dynamic propagation environments, it is essential to consider the transmitter antenna's radiation pattern and polarization characteristics. The antenna composite gain \(g^{\mathrm{A}}\) encapsulates the overall effect of these antenna properties. Specifically, the amplitude of \(g^{\mathrm{A}}\) reflects the relative antenna gain for a given direction and polarization, while its phase component describes the antenna's phase shift. 
Combining \(g^{\mathrm{A}}\) with propagation model, the signal composite gain \(g^{\mathrm{S}}_{i}\) and the phase \(\psi_{i}\) of each sub-path is obtained as,  
\begin{subequations} 
\begin{equation}
\begin{aligned}
\begin{array}{c}
g^{\mathrm{S}}_{i}=\frac{1}{\sqrt{\sum_{i=1}^{N} P_{i}}} \sum_{i=1}^{N} g^{\mathrm{A}} \cdot e^{j\psi_{i}} \cdot \sqrt{P_{i}}
\end{array},
\end{aligned}
\end{equation}
\begin{equation}
\begin{aligned}
\begin{array}{c}
\psi_{i} =\frac{2 \pi}{\lambda} \cdot f\left(d_{i}, \theta_{i}, \phi _{i}\right)
\end{array},
\end{aligned}
\end{equation}
\end{subequations}

\noindent where the phase is influenced by the combined effects of the path length \(d_{i}\), the elevation angle \(\theta_{i}\), and azimuth angle \(\phi_{i}\) , with results evaluated at the wavelength \(\lambda\) level. For the LOS path, \(d_{i}\) is the direct path length, while for the NLOS path, \(d_{i}\) is the total length after considering scattering.

\subsection{Quasi deterministic channel parameters}

To accurately represent both micro and macro channel characteristics, large-scale fading parameters and small-scale fading parameters need to be integrated, converting into channel coefficients and delay matrices that encompass multi-path effects and Doppler shifts.
The total propagation delay \(\tau_{i}\) of each sub-path is calculated based on the geometric model as,
\begin{equation}
\begin{aligned}
\tau_{i} & =\frac{1}{c} \cdot \sum_{n=1}^{n_{\mathrm{s}}} d_{i}^{n},
\end{aligned}
\end{equation}

\noindent where \(n_{\mathrm{s}}\) is the number of scatterers, $d_{i}^{n}$ is the distance between scatterers, and \(c\) is the speed of light.
The channel response coefficient \(g_{i}\) of each sub-path is obtained as,

\begin{equation}
\begin{aligned}
g_{i}= L_{\mathrm{P}} \cdot L_{\mathrm{I}} \cdot g^{\mathrm{S}}_{i},
\end{aligned}
\end{equation}

\noindent where \(L_{\mathrm{I}}\) is the combined influence of large-scale parameters. As vehicles move, the path length and angle of arrival of each sub-path vary, causing changes in \(\psi_{i}\) and \(\tau_{i}\). The rapid change of angle of arrival caused by vehicle movement manifests as the Doppler frequency shift. Therefore, the quasi deterministic channel model can consider both multi-path and Doppler effects with high spatial-temporal continuity and resolution for high-mobility vehicular networks.

\section{OTFS and delay-Doppler domain channel}
In this section, we present the mathematical formulations of OTFS modulation, along with the corresponding DD domain channel model. Then, based on the DD domain channel representation, we establish the DD domain channel prediction framework and formulate the channel prediction problem as the time series DD domain parameters prediction task.

\subsection{OTFS modulation}
OTFS modulation is an advanced scheme that maps transmitted symbols in the DD domain, allowing each symbol to experience a nearly constant channel gain even in environments with high Doppler shifts. 
By spreading each DD-domain symbol across the entire TF domain, OTFS fully exploits the diversity inherent in vehicular channels.
Consequently, this modulation technique transforms time-varying multi-path vehicular channels into sparse and stable DD-domain representations, thereby enhancing robustness and reliability of vehicular communications in dynamic wireless environments.

Let $M$ denotes the number of delay bins (subcarriers) and $N$ denotes the number of Doppler bins (time slots), and $T$ denotes the symbol duration and $\Delta f = 1/T$ denoting the subcarrier spacing. Firstly, a set of quadrature amplitude modulation (QAM) symbols $\mathbf{X}_{\text{DD}}[l,k] \in \mathbb{C}^{M \times N}$, where $l = 1, \ldots, M$ and $k = 1, \ldots, N$, is arranged on a two-dimensional DD domain grid.
This DD domain symbol matrix undergoes a transformation into the TF domain, $\mathbf{X_{TF}[\mathit{m,n}]}\in\mathbb{C^{\mathit{M}\times\mathit{N}}}$, where
 $m=1,\ldots,M$ and $n=1,\ldots,N$, by utilizing the inverse
symplectic finite Fourier transform (ISFFT) as,
\begin{equation}
\mathbf{X_{TF}}[m,n]=\frac{1}{\sqrt{MN}}\sum_{k=1}^{N}\sum_{l=1}^{M}\mathbf{\mathbf{\mathbf{X_{DD}}}}[l,k]e^{j2\pi\left(\frac{nk}{N}-\frac{ml}{M}\right)}.
\end{equation}
Subsequently, the Heisenberg transform is applied to $\mathbf{X_{TF}}$ based on the pulse waveform $g_{\mathrm{tx}}(t)$ to generate the time domain transmit signal $s(t)$ as,
\begin{equation}
  s(t)=\sum_{n=1}^{N}\sum_{m=1}^{M}\mathbf{X_{TF}}[m,n]g_{\mathrm{tx}}(t-nT)e^{j2\pi m\Delta f(t-nT)}.
\end{equation}
After passing the vehicular channel, the time-domain received signal $r(t)$ at the receiver is converted back to the TF domain matrix $\mathbf{Y_{TF} \mathit{m,n}]}\in\mathbb{C^{\mathit{M}\times\mathit{N}}}$
through the Wigner transform as,
\begin{equation}
\mathbf{Y_{TF}}[m,n]=\int g_{\mathrm{rx}}^{*}(t-nT)r(t)e^{-j2\pi m\Delta f(t-nT)}dt,
\end{equation}
where $g_{\mathrm{rx}}(t)$ is the matched pulse waveform and superscript $\ast$ denotes the conjugation. The received DD domain symbol matrix, $\mathbf{Y_{DD}[\mathit{l,k}]\in\mathbb{C^{\mathit{M}\times\mathit{N}
}}}$,
is then derived using the symplectic finite Fourier transform (SFFT) as,
\begin{equation}
\mathbf{Y_{DD}}[l,k]=\frac{1}{\sqrt{MN}}\sum_{n=1}^{N}\sum_{m=1}^{M}\mathbf{Y_{TF}}[m,n]e^{-j2\pi\left(\frac{nk}{N}-\frac{ml}{M}\right)}.
\end{equation}

\subsection{Delay-Doppler domain channel}

\begin{figure}[t]
\centerline{\includegraphics[width=6in, keepaspectratio]{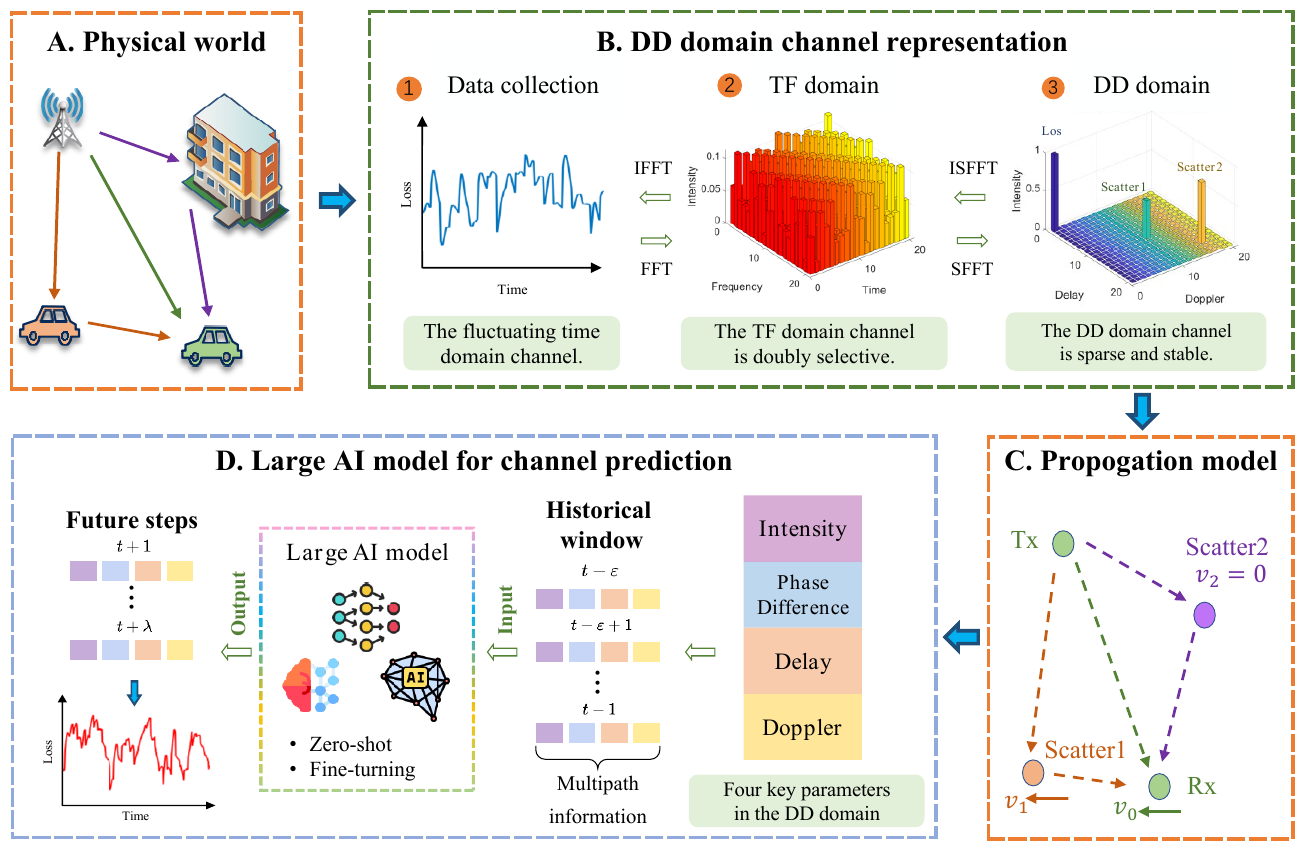}}
\caption{Delay-Doppler domain channel prediction with the large AI model.}
\label{PredictionFramework}
\end{figure}

Channel prediction in vehicular networks is particularly challenging due to the high mobility of vehicles and the complexity of the propagation environment, which results in rapid and intricate variations in the wireless channel. In the TF domain, each time and frequency point must be characterized, leading to an extensive number of channel parameters. Moreover, TF domain channels exhibit rapid fluctuations owing to the combined effects of multi-path propagation and Doppler shifts. This vast, rapidly varying parameter set significantly diminishes the predictability of the channel, thereby reducing prediction accuracy.
Transforming the channel representation to the DD domain addresses these challenges effectively by providing an intuitive, sparse, and stable channel representation.

With high time resolution channel response coefficient \(g_{i}\) and propagation delay \(\tau_{i}\) provided by the quasi deterministic channel model, we can calculate each sub-path’s Doppler shift from the phase changes in \(g(i,t)\) across neighboring snapshots. Specifically, the Doppler shift \(\nu(i)\) can be obtained as,
\begin{equation}
\begin{aligned}
\nu(i)=\frac{\Delta \arg (g(i, t_0))}{2 \pi \Delta t} ,
\end{aligned}
\end{equation}

\noindent where \(\Delta \arg (g(i, t_0))\) is the phase difference between two neighboring snapshots at time $t_0$, and \(\Delta t\) is the time interval between them. 
Assuming that within the short duration of a data frame, the relative motion between transmitter and receiver is approximately constant, the delay and Doppler shift of each sub-path remain constant. 
Then, the time-continuous gain of each sub-path \({h_i}(\tau_{i}, t)\) can be approximately expressed as, 

\begin{equation}
\begin{aligned}
{h_i}(\tau_{i}, t)=g\left(i, t_{0}\right) \cdot e^{j 2 \pi \nu(i)\left(t-\tau_{i}\right)},
\end{aligned}
\end{equation}

\noindent where \(g(i,t_{0})\) is the initial channel coefficient at time \(t_{0}\). This expression captures the Doppler-induced phase variation over time. In general, the channel with multi-path fading and Doppler shift can be modeled as a linear time-invariant system, so the input-output relationship of this channel can be obtained as,
\begin{equation}
\begin{aligned}
r(t)=
\sum_{i=1}^{N_{\mathrm{P}}} {h_i}(\tau_{i}, t) s\left(t-\tau_{i}\right)
= \sum_{i=1}^{N_{\mathrm{P}}} {g\left(i, t_{0}\right) \cdot e^{j 2 \pi \nu(i)\left(t-\tau_{i}\right)}} s\left(t-\tau_{i}\right),
\end{aligned}
\end{equation}

\noindent where $r(t)$ is the received signal and $s(t)$ is the transmitted signal, $N_{\mathrm{P}}$ is the total number of sub-paths.

Traditional orthogonal frequency division multiplexing (OFDM) works in the TF domain and therefore uses the TF domain channel representation. By taking a Fourier transform, the TF domain channel impulse response can be obtained as,
\begin{equation}
\begin{aligned}
H(f, t) = \sum_{i=1}^{N_{\mathrm{P}}} g(i, t_{0}) e^{-j 2 \pi \nu(i) \tau_i} e^{-j 2 \pi (f \tau_i - \nu(i) t)},
\end{aligned}
\end{equation}

\noindent where $f$ is frequency. For the case of the static channel where Doppler shift \(\nu(i)\) equals 0, $H(f, t)$ is reduced to time-independent frequency response $H(f)$, which is known as frequency-fading channel. In this case, the channel prediction only needs to predict channel response at different frequencies. However, the high mobility of vehicles make the channel time-variant, introducing the doubly-selective fading and making every element in $H(f, t)$ becomes different. This significantly increases the number of parameters that need to be processed for channel prediction. Therefore, it will be very complicated to make predictions for the high-mobility vehicular channel in the TF domain.

In the OTFS system, the delay-Doppler channel response is expressed in terms of a group of sub-paths with different gain, delay, and Doppler as,

\begin{equation}
\begin{aligned}
h(\tau, \nu)=
g\left(i, t_{0}\right) \cdot e^{-j 2 \pi \nu(i) \tau_{i}}
\delta\left(\tau-\tau_{i}\right) \delta\left(\nu-\nu_{i}\right),
\end{aligned}
\end{equation}

\noindent where \(\delta(\cdot)\) is the Dirac delta function. 
This formulation offers a sparse and quasi-time-invariant channel representation in the DD domain channel, closely reflecting the physical properties of the environment. This is a sparse wireless channel representation in the DD domain for the limited number of sub-paths. Therefore, the received signal $r(t)$ can be expressed as,

\begin{equation}
\begin{aligned}
r(t) = \iint h(\tau, \nu) e^{j 2 \pi \nu t} s(t - \tau) \, d\nu d\tau.
\end{aligned}
\end{equation}

Finally, we obtain the DD domain channel representation for the high-mobility scenario, in which the wireless channel at a given time snapshot $t_0$ can be completely represented by parameters including the channel response coefficient $g(i,t_{0})$, delay spread $\tau_{i}$, and Doppler shift $\nu_{i}$ of each sub-path. The DD domain channel representation with these parameters is denoted as,
\begin{equation}
\begin{aligned}
H_{t_0}^{\mathrm{DD}}=\{ g(i,t_{0}),\tau_{i},\nu_{i}\},  i=1,\ldots,N_{\mathrm{P}}.
\end{aligned}
\end{equation}

\noindent This is an intuitive, sparse, and stable channel representation, which is particularly beneficial for channel prediction in vehicular networks.

\subsection{Channel prediction problem formulation}

With the DD domain channel representation, the problem of vehicular network channel prediction becomes a task of predicting the set of the DD domain channel parameters.
We leverage the neural network for channel prediction for its powerful time series prediction capability. However, the inputs to neural networks usually require real numbers, but the channel response coefficients are complex, so it is necessary to represent the channel response coefficients as real numbers. Since the intensity of the channel response coefficients varies slowly with time and its phase varies rapidly and periodically with time at a relatively constant rate due to the Doppler effect, we express the channel response coefficients based on the form of polar co-ordinates by expressing them as intensity $|g(i,t_{0})|$ and phase differences $\Delta \varphi_{i,t_{0}}$. To satisfy the real number requirement of the neural network, the DD domain channel parameters are reformulated as,

\begin{equation}
\begin{aligned}
 H_{t_0}^{\mathrm{DD}}=\{|g(i,t_{0})|, \Delta \varphi_{i,t_{0}}, \tau_{i}, \nu_{i}\}, \mathrm{for} \enspace  i = 1, \ldots, N_{\mathrm{P}}.
\end{aligned}
\end{equation}

As shown in Fig. \ref{PredictionFramework}, the task of the DD domain channel prediction framework can be formalized as follows. Given the historical and current DD domain channel parameters of $ \varepsilon $ time snapshots, represented as  \(\left\{ H_{t-\varepsilon+1}^{\mathrm{DD}}, H_{t-\varepsilon+2}^{\mathrm{DD}}, \ldots, H_{t}^{\mathrm{DD}} \right\} \), the goal of channel prediction is to accurately forecast the time series DD domain channel parameters for the subsequent \( \lambda \) steps, denoted as \( \left\{ H_{t+1}^{\mathrm{DD}}, H_{t+2}^{\mathrm{DD}}, \ldots, H_{t+\lambda}^{\mathrm{DD}} \right\} \). Therefore, the core of the DD domain vehicular channel prediction task is to build the mapping function $\mathcal{F}(\cdot)$,
\begin{equation}
\{H_{t+1}^{DD},H_{t+2}^{DD},...,H_{t+\lambda}^{DD}\}=\mathcal{F}(\{H_{t-\varepsilon}^{DD},H_{t-\varepsilon+1}^{DD},...,H_{t-1}^{DD}\}),
\end{equation}
which aims to leverage temporal correlations within historical time series DD domain channel parameters to anticipate future channel conditions.

\section{Large AI model for channel prediction}

\begin{figure}[t]
\centerline{\includegraphics[width=6in, keepaspectratio]{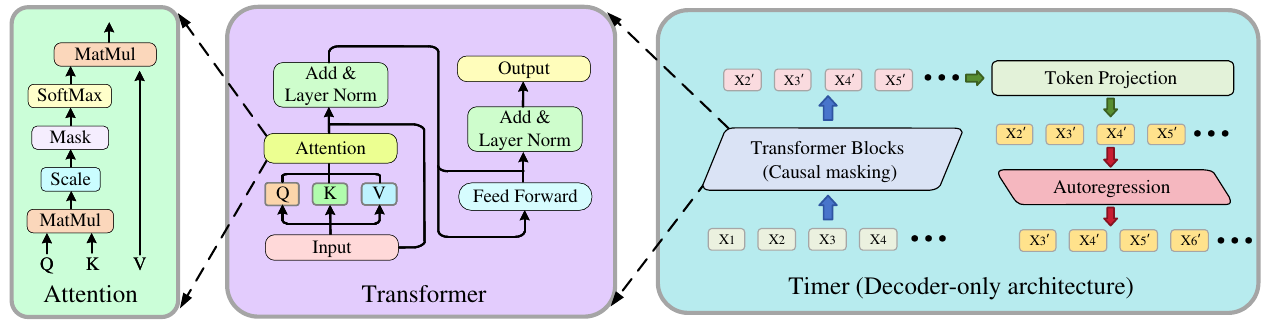}}
\caption{Large AI model with Transformer as backbone.}
\label{fig:timer}
\end{figure}

In recent years, the large AI model has become a powerful tool for processing and forecasting time series by leveraging large amounts of data and sophisticated architecture. In this section, the transformer architecture is firstly presented, serving as the backbone of the large AI model. Then we introduce the time series large AI model used for channel prediction.

\subsection{Transformer backbone}

Transformers have become the backbone of the large AI model due to their outstanding performance and efficiency. Their architecture, featuring self-attention mechanisms and parallel processing capabilities, excels at capturing dependencies and correlations in data, leading to outstanding results in various applications. These attributes are particularly advantageous for channel prediction tasks, where understanding temporal correlations and dependencies is crucial.

The core feature of the transformer architecture is the self-attention mechanism, which enables the model to consider all positions in the input sequence simultaneously. Given a sequential input $I$, the query $Q$, key $K$, value $V$ are computed as,
\begin{equation}
\begin{cases}
Q=IW^{Q},\\
K=IW^{K},\\
V=IW^{V},
\end{cases}
\end{equation}
where $W^{Q}$, $W^{K}$, $W^{V}$ are learnable weight matrices. The self-attention mechanism is then calculated as,
\begin{equation}
\mathrm{Attention}(Q,K,V)=\mathrm{softmax}(\frac{QK^{\mathrm{T}}}{\sqrt{d_{k}}})V,
\end{equation}
where $d_{k}$ is the dimension of the key vectors used as a scaling factor, $\mathrm{softmax}(\cdot)$ is
an activation function.
Building upon the self-attention mechanism, the transformer implements multi-head attention. It allows the model to learn information from different representation subspaces, thereby enriching its ability to capture a wider range of dependencies. The multi-head attention is expressed as,
\begin{equation}
\mathrm{MultiHead}(Q,K,V)=\mathrm{Concat}(head_{c}|c=1,2,...,C)W^{O},
\end{equation}
\begin{equation}
head_{c}=\mathrm{Attention}(QW_{c}^{Q},KW_{c}^{K},VW_{c}^{V}),
\end{equation}
\noindent where $C$ is the number of the parallel
attention heads, $W_{c}^{Q}$, $W_{c}^{K}$, $W_{c}^{V}$ are the weight matrices for the c-th head, $W^{O}$ is a learnable projection matrix.

\subsection{Time series large AI model}
We employ a large AI model named Timer for channel prediction. As shown in Fig. \ref{fig:timer}, Timer, short for time series transformer, is a large pre-trained model specifically designed for processing time-series data, utilizing the Transformer architecture and generative pre-training to provide a unified approach to various time-series tasks \cite{Timer}. For specific down-stream applications, Timer can be fine-tuned on task-specific data, allowing it to adapt to the nuances of the target task while leveraging knowledge gained during pre-training. Pre-trained time series large AI model offers the advantage of low-cost and rapid deployment of channel prediction through zero-shot learning, where the model performs tasks on channel data it has never encountered by leveraging learned knowledge. Additionally, fine-tuning further enhances channel prediction accuracy by retraining the pre-trained model with specific channel data.

Assume a univariate variate time series sequence $X=\{x_{1},...,x_{UJ}\}$ of unified context length $UJ$ is the input to the Timer model. Timer tokenizes the input data into consecutive time segments that encompass the series variations as,
\begin{equation}
j_{u}=\{x_{(u-1)J+1},...,x_{uJ}\}\in\mathbb{R}^{J},u=1,...,U,
\end{equation}
where $J$ is the length of each time segment, $U$ is the number
of tokens of the input data. Timer employs a decoder-only Transformer with 
dimension $D$ and $L$ layers, and applies generative pre-training
on $U$ tokens of the sequence. The computations are as follows,

\begin{equation}
z_{u}^{0}=W_{e}j_{u}+TE_{u},u=1,...,U,
\end{equation}

\begin{equation}
Z^{l}=\mathrm{TransformerBlock}(Z^{l-1}),l=1,...,L,
\end{equation}

\begin{equation}
\{\hat{j}_{u+1}\}=Z^{L}W_{d},u=1,...,U,
\end{equation}
where $W_{e}$, $W_{d}$$\in\mathbb{R}^{D\times J}$ are the encoding and decoding matrices for token embeddings in $H=\{h_{u}\}\in\mathbb{R}^{U\times D}$, respectively. $TE_{u}$ is the optional time stamp embedding. Using the causal
attention within the decoder-only transformer, the model autoregressively generates $\hat{j}_{u+1}$ as the prediction for the next token given $j_{u}$. The pre-training objective is formulated to minimize the mean squared error between the predicted and actual tokens,

\begin{equation}
\mathcal{L}=\frac{1}{UJ}\parallel j_{u}-\hat{j}_{u}\parallel_{2}^{2},u=2,...,U.
\end{equation}
This loss function provides token-wise supervisory signals, ensuring that the generated tokens at each position are independently supervised. By leveraging the Transformer architecture's ability to capture long-range dependencies through self-attention, Timer effectively models the temporal correlations in time series data, enhancing its predictive capabilities.

By pre-training on large and diverse time series datasets, Timer learns to recognize general patterns and trends across different sequences, automatically extracting important temporal features and understanding underlying temporal relationships. This pre-training endows Timer with exceptional zero-shot learning capability for processing time series data. Since DD domain channel parameters vary regularly over time, forming a time series, they can be effectively predicted using Timer. Leveraging its learned temporal features, Timer can generalize to unseen channel data, enabling it to perform channel prediction tasks without prior training on specific channel data. Furthermore, fine-tuning Timer with collected channel data can significantly improve its performance, demonstrating its effectiveness in vehicular network prediction. Therefore, the zero-shot capability of large AI model like Timer enables low-cost channel prediction deployment, while fine-tuning further enhances prediction accuracy.

\section{Simulation results}
In this section, we present simulation results to evaluate the effectiveness of our proposed approach. The experimental setups for the vehicular network scenario and the large AI model are first given. Then, we analyze the prediction accuracy in terms of both DD domain parameters and total path loss, respectively.

\subsection{Experimental settings}

We consider vehicle-to-infrastructure (V2I) channels in a 500-meter-long highway scenario, where a base station with a single antenna is located 100 meters from the road center. The road has two lanes in each direction, spaced 5 meters apart. Vehicles, equipped with single antennas, travel at speeds two different classic speeds including 60 and 120 kilometers per hour. For each pair of side-by-side vehicles traveling in the same direction, one is designated for training and the other for testing. The vehicular channels are generated using the quasi-deterministic radio channel generator (QuaDRiGa), considering both LOS and NLOS conditions \cite{QuaDriGa}. The channels are sampled with a temporal resolution of 0.5 milliseconds. 
The pre-processing of DD domain channel parameters is performed using z-score normalization. Each of the four parameter types is normalized independently, converting data with varying scales and distributions into a common scale with a mean of 0 and a standard deviation of 1.
In addition, the performance of channel prediction is evaluated using a weighted mean absolute error (MAE), where the weighting is determined by the intensity of the different sub-paths. The GPU used in the experiment is the NVIDIA GeForce RTX 4090D.

For the time series large AI model named Timer, the model dimension is configured to 1024 and the feed forward dimension is 2048, while it employs 8 heads within the multi-head attention mechanism \cite{Timer}. During the fine-tuning process, the learning rate is $1 \times 10^{-5}$ and the loss function is MAE. Another large AI model named Time-MoE (mixture of experts) is also used for zero-shot prediction \cite{TimeMoE}. Two classical models including LSTM and GRU are used to predict the channel after training. The input time series length is 20 and the output time series length is 10, indicating that the model predicts future channel conditions based on 10 milliseconds of historical data to forecast the subsequent 5 milliseconds.

\subsection{Prediction accuracy of delay-Doppler domain parameter}

\begin{table*}[!t]
\centering
\footnotesize
\caption{Comparison of prediction errors for different velocities in LOS channel scenario.}
\label{tab:results-1}
\begin{tabular}{|c|c|cccc|}
\hline 
\multirow{1}{*}{Model} & \multirow{1}{*}{Speed(km/h)} & Delay(s) & Doppler shift(Hz) & Intensity(dB) & Phase difference(rad)\tabularnewline
\hline 
\multirow{2}{*}{LSTM} & 60 & 1.06E-09 & 0.1239 & 0.0647 & 0.0041\tabularnewline
 & 120 & 2.30E-09 & 0.3870 & 0.1010 & 0.0124\tabularnewline
\hline 
\multirow{2}{*}{GRU} & 60 & 3.44E-10 & 0.0959 & 0.0559 & 0.0031\tabularnewline
 & 120 & 8.96E-10 & 0.2774 & 0.0877 & 0.0092\tabularnewline
\hline 
\multirow{2}{*}{Time-MoE (zero-shot)} & 60 & 4.16E-10 & 0.0202 & 0.0438 & 0.0013\tabularnewline
 & 120 & 7.05E-10 & \textbf{0.0504} & 0.2488 & 0.0084\tabularnewline
\hline 
\multirow{2}{*}{Timer (zero-shot)} & 60 & 1.64E-11 & 0.0342 & 0.0202 & 0.0010\tabularnewline
 & 120 & 5.67E-11 & 0.2041 & 0.0986 & 0.0062\tabularnewline
\hline 
\multirow{2}{*}{Timer (fine-tuning)} & 60 & \textbf{6.06E-12} & \textbf{0.0176} & \textbf{0.0081} & \textbf{0.0005}\tabularnewline
 & 120 & \textbf{2.30E-11} & {0.1215} & \textbf{0.0519} & \textbf{0.0036}\tabularnewline
\hline 
\end{tabular}
\end{table*}

\begin{table*}[!t]
\centering
\footnotesize
\caption{Comparison of prediction errors for different velocities in NLOS channel scenario.}
\label{tab:results-2} 
\begin{tabular}{|c|c|cccc|}
\hline 
\multirow{1}{*}{Model} & \multirow{1}{*}{Speed(km/h)} & Delay(s) & Doppler shift(Hz) & Intensity(dB) & Phase difference(rad)\tabularnewline
\hline 
\multirow{2}{*}{LSTM} & 60 & 4.36E-09 & 0.4370 & 0.2253 & 0.0128\tabularnewline
 & 120 & 3.68E-09 & 1.0428 & 0.2600 & 0.0330\tabularnewline
\hline 
\multirow{2}{*}{GRU} & 60 & 1.79E-09 & 0.4117 & 0.2280 & 0.0124\tabularnewline
 & 120 & 1.57E-09 & 0.9375 & 0.2525 & 0.0301\tabularnewline
\hline 
\multirow{2}{*}{Time-MoE (zero-shot)} & 60 & 1.19E-09 & 0.0819 & 0.1820 & 0.0054\tabularnewline
 & 120 & 1.38E-09 & \textbf{0.1360} & 0.5234 & 0.0178\tabularnewline
\hline 
\multirow{2}{*}{Timer (zero-shot)} & 60 & 7.26E-11 & 0.1353 & 0.0393 & 0.0043\tabularnewline
 & 120 & 1.43E-10 & 0.5045 & 0.1218 & 0.0160\tabularnewline
\hline 
\multirow{2}{*}{Timer (fine-tuning)} & 60 & \textbf{3.14E-11} & \textbf{0.0809} & \textbf{0.0226} & \textbf{0.0026}\tabularnewline
 & 120 & \textbf{6.33E-11} & {0.3212} & \textbf{0.0791} & \textbf{0.0104}\tabularnewline
\hline 
\end{tabular}
\end{table*}

Table \ref{tab:results-1} and Table \ref{tab:results-2} present a comprehensive analysis of different models' effectiveness in predicting channel conditions in vehicular networks under both LOS and NLOS scenarios. The results highlight the significant advantage of the fine-tuned Timer model, which consistently outperforms classical models and zero-shot large AI models in most metrics with vehicle speeds of 60 km/h and 120 km/h. In LOS conditions, the fine-tuned Timer demonstrates its ability to capture channel dynamics effectively, even as vehicle speed increases, where traditional models like LSTM and GRU struggle to adapt. In NLOS conditions, where channel characteristics become even more complex due to multi-path propagation, the fine-tuned Timer maintains its advantage, highlighting its robustness in handling the severe Doppler shifts and multi-path effects characteristic of NLOS environments. The results also underscore the value of the zero-shot capability inherent in large AI models like Timer and Time-MoE, which enables reasonably accurate channel predictions without specific training, thus facilitating low-cost deployment in new scenarios. Moreover, fine-tuning further refines Timer’s predictive accuracy and enables it to adapt to the unique characteristics of vehicular channels, emphasizing the additional accuracy gained through fine-tuning for task-specific data. In comparison, classical models like LSTM and GRU exhibit considerably higher prediction errors across both LOS and NLOS scenarios, especially as vehicle speed increases, indicating their limitations in capturing the rapid and complex dynamics of vehicular channels.

\begin{figure*}[!t]
\centering
\subfloat[LOS channel scenario.]{\includegraphics[width=3in]{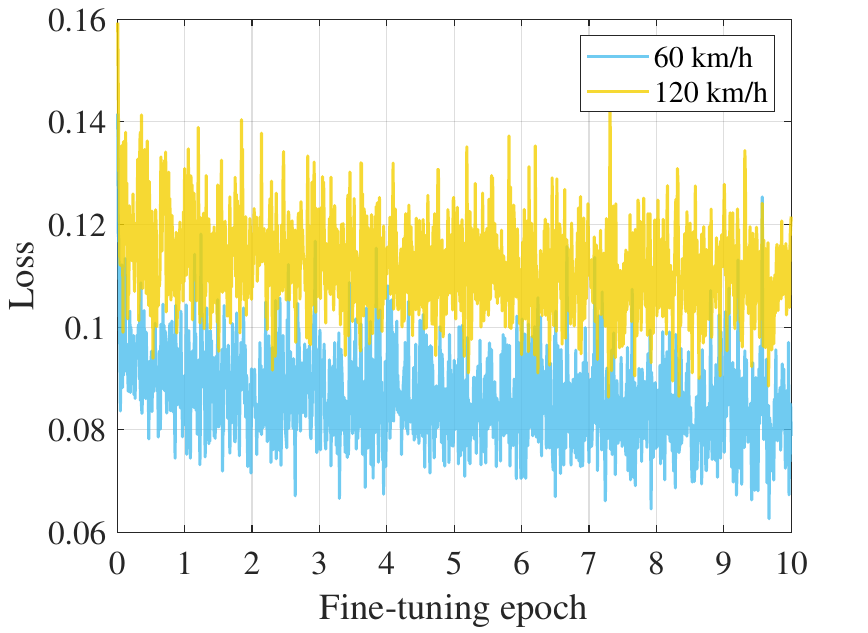}\label{fig:los_loss}}
\subfloat[NLOS channel scenario.]{\includegraphics[width=3in]{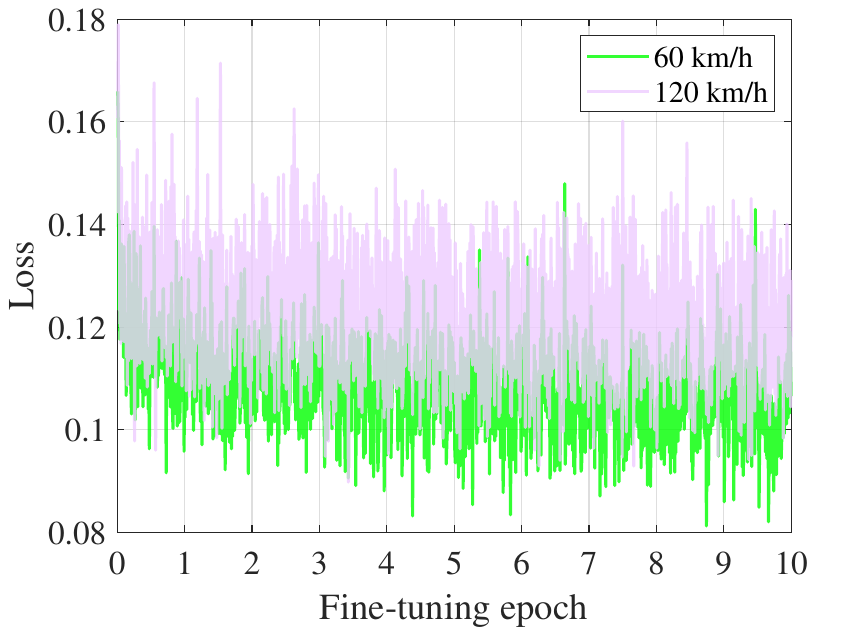}\label{fig:nlos_loss}}
\caption{Loss curves of fine-tuning for the large AI model.}
\label{fig:loss}
\end{figure*}

\begin{figure*}[!t]
\centering\subfloat[Delay.]{\includegraphics[width=1.5in]{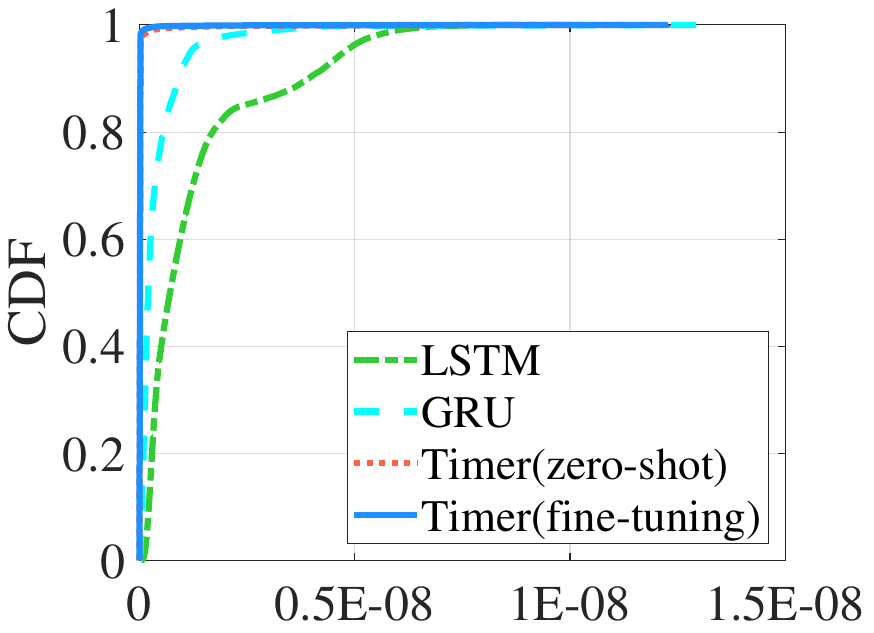}\label{fig:cdf_delay}

}\subfloat[Doppler shift.]{\includegraphics[width=1.5in]{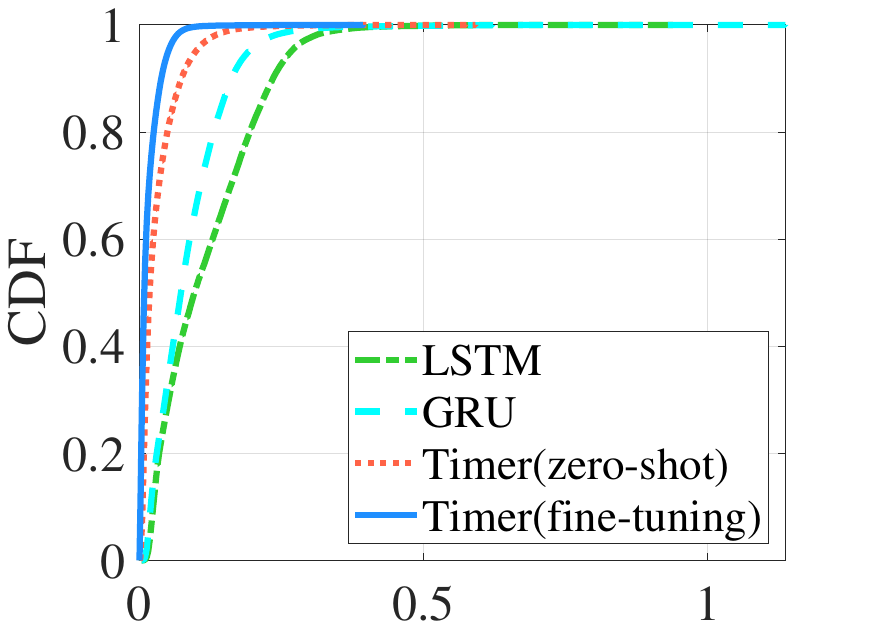}\label{fig:cdf_doppler}

}\subfloat[Intensity.]{\includegraphics[width=1.5in]{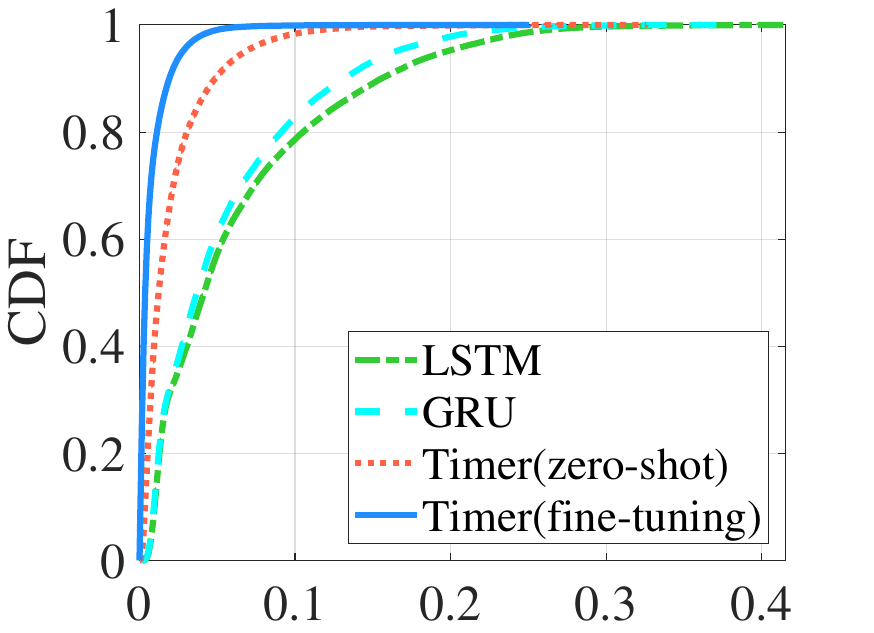}\label{fig:cdf_intensity}

}\subfloat[Phase difference.]{\includegraphics[width=1.5in]{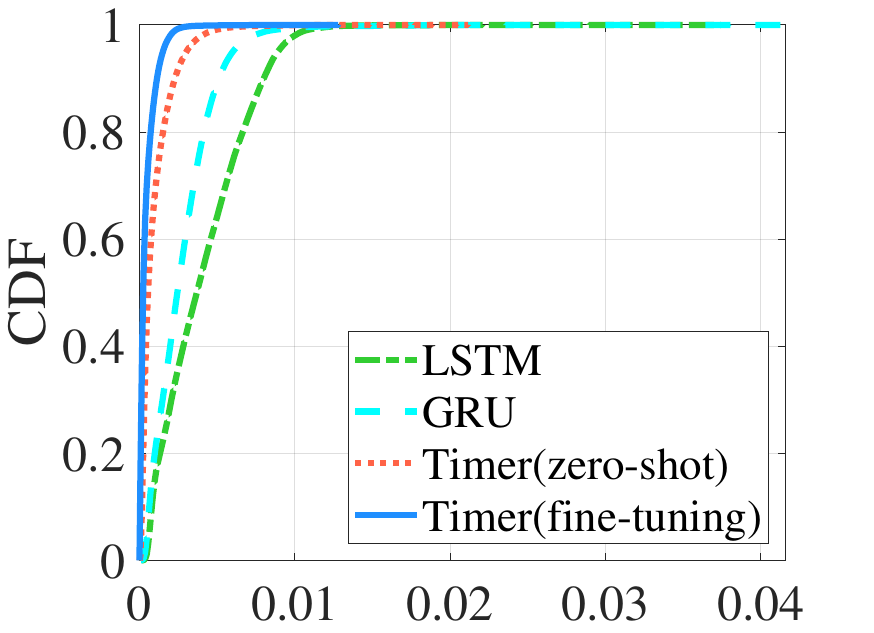}\label{fig:cdf_phase}

}\caption{CDF of prediction errors for different delay-Doppler domain parameters.}
\label{fig:cdf}
\end{figure*}

Figure \ref{fig:loss} shows the loss curves of Timer during the fine-tuning process under LOS and NLOS channel conditions at vehicle speeds of 60 km/h and 120 km/h. In the LOS scenario shown in Fig. \ref{fig:los_loss}, the loss curve for 120 km/h consistently remains higher than that for 60 km/h. This indicates that as vehicle speed increases, the complexity of the channel characteristics in the LOS scenario also increases, resulting in higher loss values. In contrast, in the NLOS scenario shown in Fig. \ref{fig:nlos_loss}, the loss curves for both speeds overlap more substantially, indicating that NLOS conditions are already sufficiently complex such that further increases in speed do not markedly exacerbate this complexity. Across both scenarios, all curves stabilize after approximately 1000 fine-tuning steps, illustrating Timer’s rapid convergence and efficiency in adapting to varying channel conditions. This rapid convergence highlights Timer's efficiency in adapting to different vehicular channel prediction scenario, requiring only a relatively short fine-tuning period to perform effectively in specific tasks. This adaptation is beneficial in dynamic vehicular environments, where the large AI model can be accurately, flexibly and cost-effectively deployed.

Fig. \ref{fig:cdf} shows the cumulative distribution function (CDF) of prediction errors for various models across four different DD domain channel parameters in the LOS scenario at 60 km/h. The fine-tuned Timer consistently outperforms all other models, with CDF curves that reach 1 quickly for each parameter, indicating minimal prediction errors and high accuracy. Timer in zero-shot mode performs well but falls short of the fine-tuned version, highlighting the benefits of fine-tuning. In contrast, LSTM and GRU display broader CDF curves, reflecting higher prediction errors and less reliability across all DD domain channel parameters. Specifically, the fine-tuned Timer excels in capturing Doppler shift and phase difference accurately, which are critical for handling mobility-induced channel variations, while LSTM and GRU lag considerably, suggesting limitations in adapting to dynamic vehicular conditions. Overall, this analysis underscores the advantage of Timer, particularly when fine-tuned, as an accurate and robust model for channel prediction in high-mobility vehicular environments.

\subsection{Prediction accuracy of path loss}

\begin{figure*}[!t]
\centering
\subfloat[Speed at 60 km/h.]{\includegraphics[width=3in]{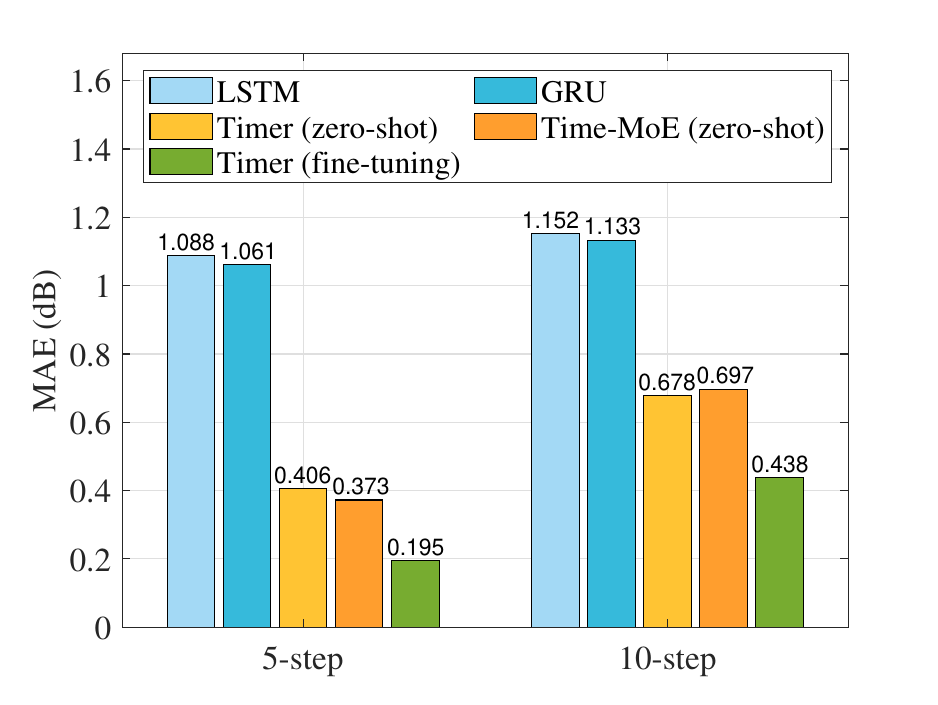}\label{fig:60_LOS}}
\subfloat[Speed at 120 km/h.]{\includegraphics[width=3in]{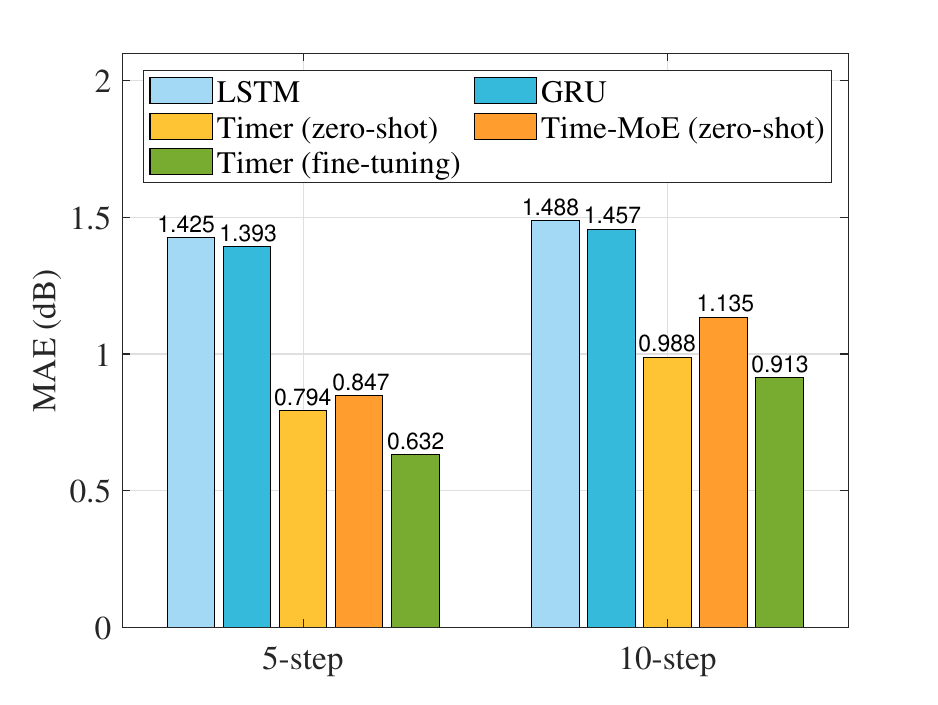}\label{fig:120_LOS}}
\caption{ Comparison of mean absolute errors in LOS channel scenario.}
\label{fig:loss_LOS}
\end{figure*}

\begin{figure*}[!t]
\centering
\subfloat[Speed at 60 km/h.]{\includegraphics[width=3in]{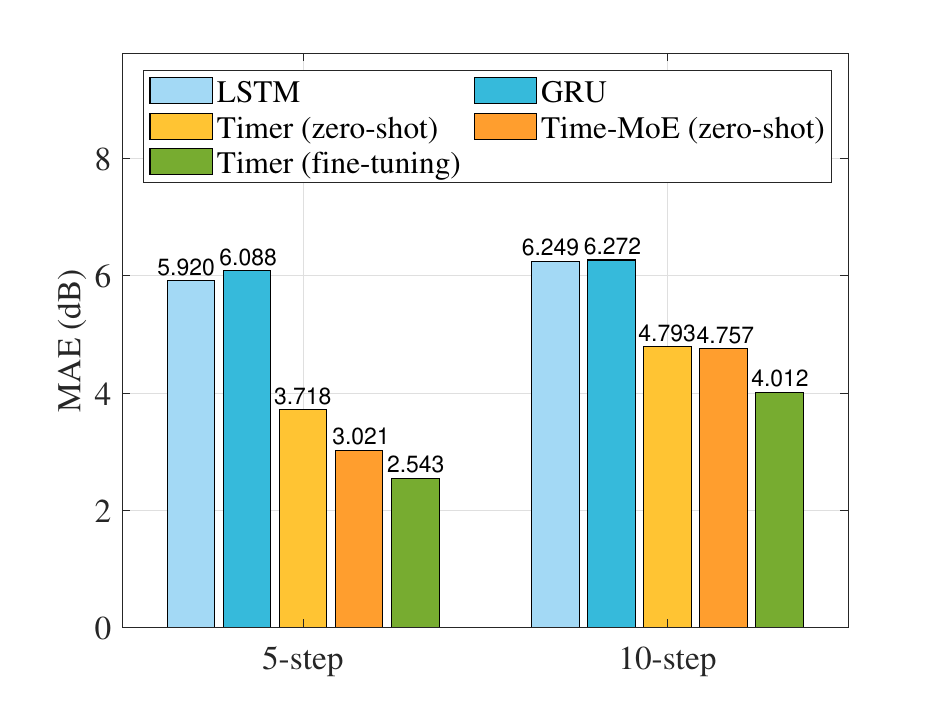}\label{fig:60_NLOS}}
\subfloat[Speed at 120 km/h.]{\includegraphics[width=3in]{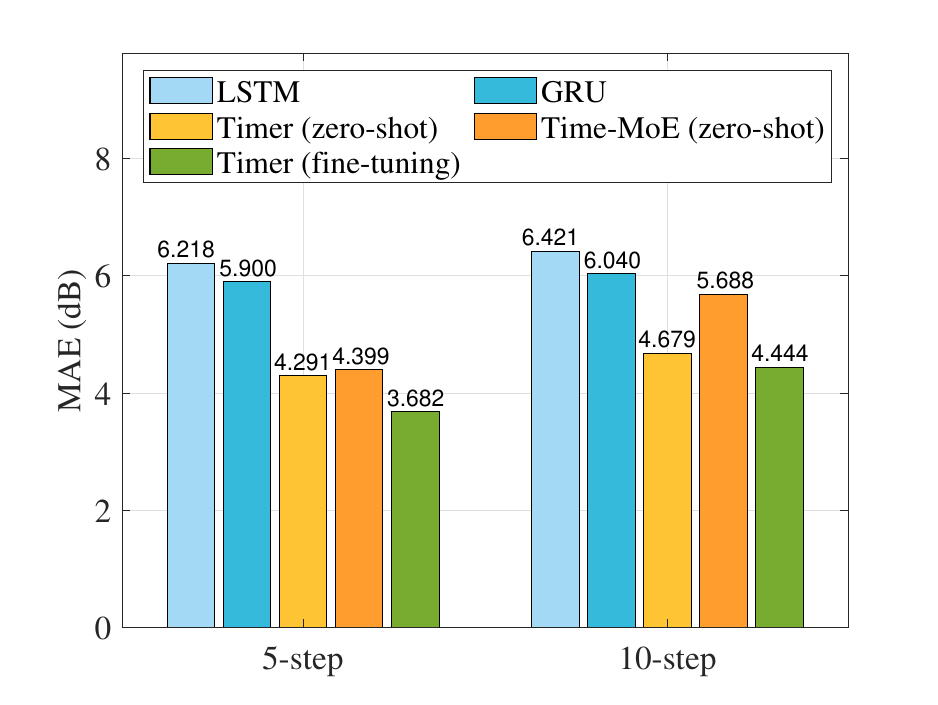}\label{fig:120_NLOS}}
\caption{Comparison of mean absolute errors in NLOS channel scenario.}
\label{fig:loss_NLOS}
\end{figure*}

Fig. \ref{fig:loss_LOS} and Fig. \ref{fig:loss_NLOS} compare the path loss prediction performance of five models across LOS and NLOS conditions at vehicle speeds of 60 km/h and 120 km/h, with prediction step length of 5-step (2.5 ms) and 10-step (5 ms). The path loss is obtained by combining the DD domain parameters at each time step. Zero-shot capability of Timer and Time-MoE also perform well, indicating the advantage of large AI models for low-cost and generalized deployment without specific training for each environment. Moreover, fine-tuning further enhances Timer’s accuracy, allowing it to capture unique channel characteristics. The fine-tuned Timer model consistently achieves the lowest MAE across all scenarios, demonstrating its adaptability and robustness in dynamic vehicular environments. Classical models, such as LSTM and GRU, show higher MAEs, particularly in NLOS conditions and at 120 km/h, highlighting their limitations in handling the complex, rapidly fluctuating channels typical of vehicular networks.
When comparing the 5-step and 10-step prediction performance, all models exhibit higher MAE at the longer prediction steps due to the increased uncertainty. Notably, the fine-tuned Timer model shows a comparatively smaller increase in MAE, indicating its superior ability to capture and extrapolate channel dynamics over longer time spans.
Overall, these results emphasize that Timer, especially when fine-tuned, offers a powerful combination of precision, adaptability, and cost-effective deployment potential for vehicular channel prediction.

\begin{figure*}[!t]
\centering
\subfloat[Path loss with small variation.]{\includegraphics[width=3in]{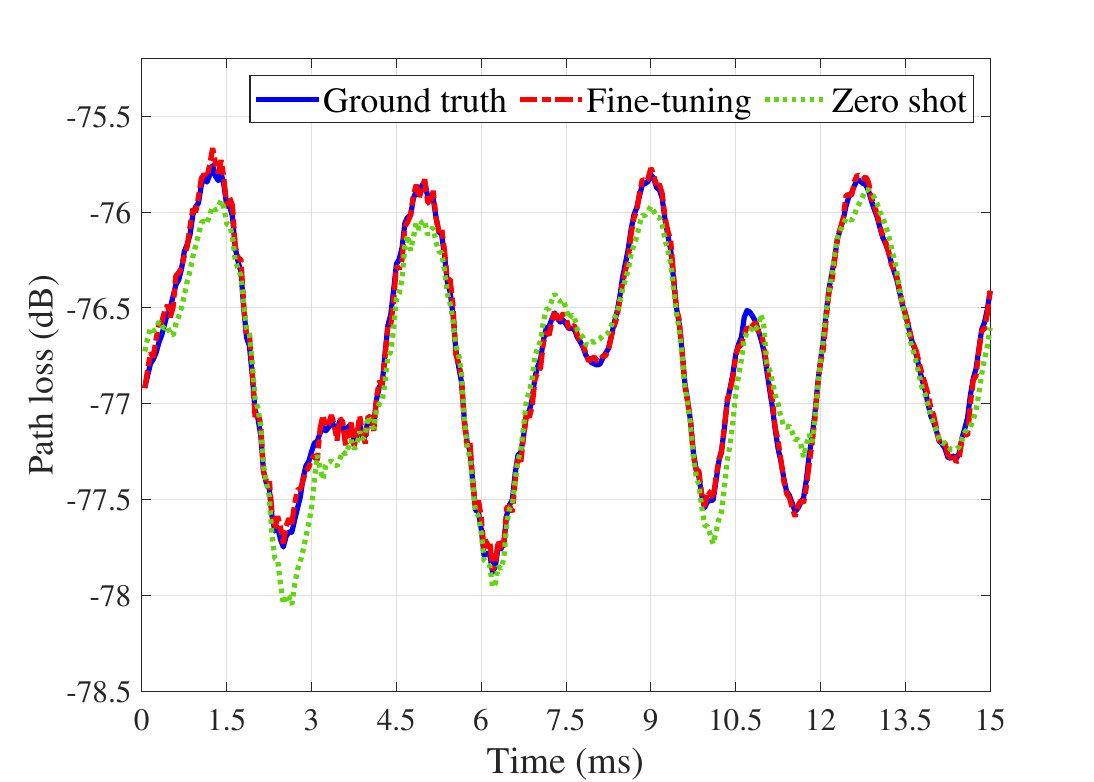}\label{fig:compare_small}}
\subfloat[Path loss with large variation.]{\includegraphics[width=3in]{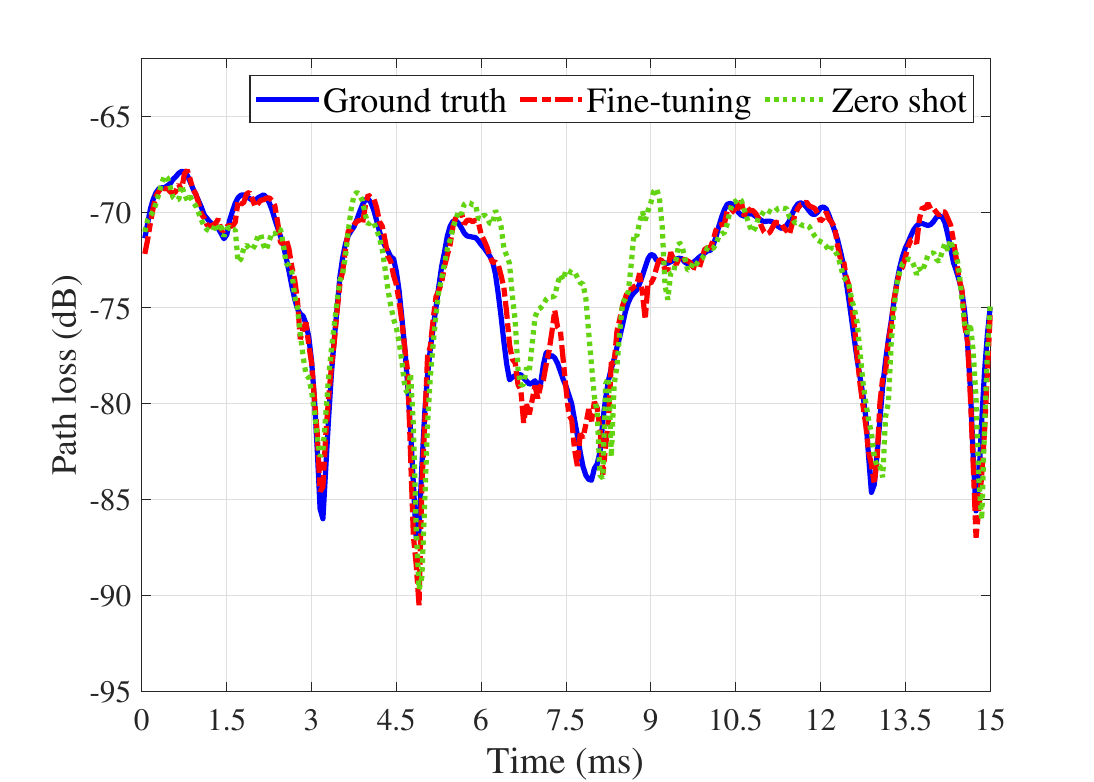}\label{fig:compare_large}}
\caption{Visualization of the predicted path loss using the large AI model.}
\label{fig:comparsion}
\end{figure*}

Fig. \ref{fig:comparsion} shows a visual comparison of predicted path loss in the LOS scenario with the vehicle speed of 60 km/h, comparing the ground truth path loss values with predictions from the fine-tuned Timer model and the zero-shot Timer model over two different 15 ms time periods: one with small path loss variation and another with large path loss variation. In the small variation case shown in Fig. \ref{fig:compare_small}, where path loss fluctuates within a 2 dB range, the fine-tuned Timer model closely follows the real path loss trajectory, demonstrating its ability to accurately capture subtle variations. The zero-shot Timer, while capturing the general trend, shows larger deviations from the true values, especially at peaks and troughs, indicating reduced accuracy without fine-tuning.
In the large variation case shown in Fig. \ref{fig:compare_large}, where path loss fluctuates by nearly 10 dB, the fine-tuned Timer again closely mirrors the real path loss, maintaining accuracy even with significant fluctuations. This consistency in performance highlights the fine-tuned model’s adaptability to both stable and highly dynamic channel conditions, a critical feature in high-mobility vehicular environments. The zero-shot Timer, however, struggles to match the rapid variations, showing more pronounced deviations from the real path loss, particularly during sharp changes. These observations underscore the advantage of fine-tuning, which enables Timer to adapt its predictions closely to real data patterns, improving accuracy and making it more reliable in dynamic conditions.

The results show the value of zero-shot capability in large AI models like Timer, allowing for cost-effective channel prediction deployment without specific training, which can be useful for vehicular networks with complex channel conditions. Meanwhile, fine-tuning significantly enhances Timer’s predictive accuracy, aligning it more closely with true path loss values, particularly in scenarios with rapid changes. This comparison highlights that while zero-shot models provide a baseline performance, fine-tuning is essential for achieving high accuracy in complex, high-variation environments.

\section{Conclusion}
In this paper, we have proposed a DD domain channel prediction framework for high-mobility vehicular networks, in which the large AI model is leveraged for the DD domain channel parameter prediction. By utilizing the DD domain, we have transformed complex vehicular channels into a set of time series parameters with high predictability. Extensive simulation results have demonstrated that the zero-shot capability of large AI model can provide low-cost channel prediction deployments, while fine-tuning can further achieve high prediction accuracy in complex vehicular networks. 
Moreover, our framework is highly applicable to a broad range of high-mobility communication systems, including unmanned aerial vehicle (UAV) and satellite-terrestrial communications.
For the future work, we will investigate link adaptation in vehicular networks aided by channel prediction.

\Acknowledgements{This work is supported in part by the National Natural Science Foundation of China (Grant Nos. 623B2052 and 62271244).}

\bibliographystyle{scis}
\bibliography{CP_Ref}

\end{document}